\documentstyle[psfig]{mn}
\newcommand{\xib}{\bar\xi}
\newcommand{\beq}{\begin{equation}}
\newcommand{\eeq}{\end{equation}}
\newcommand{\noi}{\noindent}
\newcommand{\lb}{\left(}
\newcommand{\rb}{\right)}
\date{}
\title[The effects of anti-correlation on gravitational clustering.]{The effects of anti-correlation on gravitational clustering.}
\author[Nissim Kanekar and T. Padmanabhan]
{Nissim Kanekar$^{1}$ \thanks{nissim@ncra.tifr.res.in} and 
T. Padmanabhan$^{2}$\thanks{paddy@iucaa.ernet.in}\\
$^{1}$National Centre for Radio Astrophysics,\\
 Post Bag 3, Ganeshkhind, Pune 411 007, India \\
$^{2}$Inter-University Centre for Astronomy and Astrophysics,\\
 Post Bag 4, Ganeshkhind, Pune 411 007, India }
\begin{document}
\maketitle
\begin{abstract}
We use non-linear scaling relations (NSRs) to investigate the effects arising 
from the existence of negative correlations 
on the evolution of gravitational clustering in an expanding universe. 
It turns out that such anti-correlated regions have important dynamical 
effects on {\it all} scales. In particular, the mere existence of negative 
values for the linear two-point correlation function $\xib_L$ over some range of 
scales starting from $l = L_o$, implies that the non-linear correlation function is 
bounded from above at {\it all} scales $x < L_o$. This also results in the 
relation $\xib \propto x^{-3}$, at these scales, at late times, independent of 
the original form of the correlation function. Current observations do not 
rule out the existence of negative $\xib$ for $ 200 h^{-1}$ Mpc $\la \xib \la 1000 
h^{-1}$ Mpc; the present work may thus have relevance for the real Universe. 
The only assumption made in the analysis is the {\it existence} of NSR; the 
results are independent of the form of the NSR as well as of the stable 
clustering hypothesis.
\end{abstract}
\begin{keywords}
Cosmology : theory -- dark matter, large scale structure of the Universe
\end{keywords}

\section{Introduction : The importance of being negative}

The problem of gravitational clustering in an expanding Universe
is one of the important open issues in cosmology. In the standard picture,
non-linear structures were formed through the mechanism of
gravitational instability, via the amplification of small initial density
perturbations. At early epochs, when the fluctuations are in the linear
regime, perturbation theory can be used to study the evolution. Analytical
understanding of the quasi- and non-linear regimes has unfortunately proved
a more intractable problem and much of the activity in this area has centred on
numerical simulations. It is, however, important to {\it understand} the numerical 
results; a number of attempts have hence been made to address the
subject by semi-analytic means. While this program is still in its infancy
and we are far from providing a ``first principles'' paradigm, such attempts
have nevertheless been successful in identifying at least some of the key
issues in the study of gravitational clustering. 

A key problem in this area involves understanding how power gets transferred
between different scales during the evolution of clustering. In the simplest
context, one would like to understand how an initial power spectrum, peaked
around a particular scale, evolves with time. Numerical results 
\cite{jasjeet} and a few analytic insights (especially regarding inverse cascade; 
see Padmanabhan 2000) show that power injected at a scale $L$ cascades both downwards 
(to $x < L$) and upwards (to $x>L$). To the lowest order of approximation, the Fourier
transform of the gravitational potential is described by the equation 
\cite{iran}:
\begin{eqnarray}
\label{calgxx}
\ddot \phi_{\bf k} + 4 {\dot a \over a} \dot \phi_{\bf k} &=& -{1 \over 3a^2} \int 
{d^3 \bf p \over (2 \pi)^3} \phi_{{1 \over 2} \bf k + \bf p} \phi_{{1 \over 2} 
\bf k - \bf p} {\cal G} (\bf k, \bf p) \\
{\cal G} (\bf k, \bf p) &= &{7 \over 8} k^2 + {3 \over 2} p^2 - 5 \lb {\bf k 
\cdot \bf p\over k}\rb^2 \;\; ,\nonumber 
\end{eqnarray}
which governs the growth of $\phi_{\bf k}$ due to non-linear mode coupling.
When the non-linear coupling term on the right hand side is small, we get the
standard result that $\phi_{\bf k}$ is independent of time in the linear regime.
If, however, the initial power spectrum is sharply peaked at some scale, 
the issue  becomes more subtle. Linear theory remains valid at all scales only as 
long as $\phi_{\bf k}$ is small at {\it all} scales. As soon as some scale goes 
non-linear, it is possible for the integral on the right hand side of equation 
(\ref{calgxx}) to pick up significant contributions to drive the evolution of 
$\phi_{\bf k}$ at {\it other} scales. For example, this will lead to a $k^4$ tail 
in the power spectrum of the density contrast \cite{iran}, purely 
due to non-linear coupling. This is a clear case of inverse cascade of power 
in gravitational clustering.

Power transfer also occurs to smaller scales but 
is somewhat more difficult to analyse from first principles. Simulations 
\cite{jasjeet} as well as theoretical arguments \cite{padeng} 
suggest that the evolution to smaller scales will lead to a universal form of the power 
spectrum, reminiscent of the Kolmogorov spectrum in fluid turbulence, if the initial
spectrum was sharply peaked at some scale.  \\
\noi The above results emphasise the simplicity and tractability of a  question
involving the evolution of a power spectrum which was originally peaked around
some scale. This issue is important in understanding gravitational clustering, 
in spite of the fact that currently fashionable models for structure formation 
have broad band power spectra, with $P(k) \propto k$ for small $k$. To study
the evolution of such power spectra analytically, we need to use some technique 
which is sufficiently general and yet tractable. One such approach is based on 
non-linear scaling relations (NSRs, hereafter), which provide a mapping between the linear 
and non-linear mean two-point correlation functions \cite{ham,rajpad,mjw,pad96}. 
The above mapping appears to be validated by numerical studies, which have also 
shown it to be fairly model independent. Given an initial mean correlation function, 
$\bar\xi_{\rm L}$, in the linear regime, the NSR yields the evolved mean correlation 
function, $\bar\xi$, (and thus, the power spectrum) at any later time. It also illustrates
how power is transferred between different scales during the evolution of
clustering. Finally, the limiting forms of the NSR can be derived using the
standard paradigms of scale-invariant radial collapse and stable clustering
in the quasi-linear and non-linear regimes respectively \cite{pad96}; it has
also been shown that scaling laws are a generic feature of clustering
in an arbitrary number of dimensions \cite{padkan}.

Non-linear scaling relations have thus proved a powerful tool for
understanding gravitational clustering; however, they have one major defect. The standard 
NSR \cite{ham}, by its very form, is restricted in its applicability to positive 
semi-definite correlation functions; negative values of $\bar\xi$ are not included 
in its gamut.  However, any sufficiently localised power spectrum (for example, a 
Dirac delta function), will produce negative $\bar\xi$ at some scales; in fact, as 
will be seen later, negative $\xib$ can also arise from broad band power spectra. It 
is thus interesting, from a purely mathematical perspective, to investigate the effects
of anti-correlation on the formation of structure. 

The standard NSR shows that power is transferred from large scales to
small scales during the evolution. It turns out, however, that the very existence
of negative $\bar\xi$, combined with the equation of conservation of pairs,
implies that the flow of power will {\it reverse} direction at the scale at which
$\bar\xi$ crosses zero. Further, it will be shown that the transfer of power
to small scales saturates at late times; this results in an upper bound on $\bar\xi$
at all scales smaller than the one at which $\bar\xi$ is initially negative. Finally,
the correlation function at these scales satisfies the relation $\bar\xi \propto x^{-3}$,
at late times, independent of its original form. 

None of the above features arise for initial power spectra with positive
definite correlation functions. It should be stressed, however, that these results 
are valid even if $\xib$ is only negative over a very small range of scales or 
at arbitrarily large scales; the results also do not depend on the {\it extent} 
to which $\xib$ goes negative but arise entirely due to the {\it existence} of 
negative $\xib$. They are also quite independent of the stable clustering ansatz.
 Current observations do not preclude negative values of the 
correlation function at some scales; in fact, we will show in section \ref{obs} that one 
can indeed modify the power spectrum of the universe, so as to have negative 
correlation functions over certain ranges of scales, {\it without violating any 
observational constraints}.

The plan of the paper is as follows : the mathematical framework is
set up in section \ref{sec:Math}; we also discuss the usual NSR for the positive $\xib$
regime here. Next, NSRs for other statistical indicators of clustering, such as the 
power spectrum, are considered in section \ref{nsr_compare}. We move on to 
review observational constraints on the correlation function and the power
spectrum in section \ref{obs}, demonstrating (by example) that 
it is possible to set up power spectra yielding negative correlation functions without 
violating any observational constraints. Section \ref{negxi} considers the effects of 
negative 
correlation functions on the evolution of clustering and shows that the existence of 
anti-correlated regions yields an upper bound on the correlation function at scales 
shortward of the scale at which the linear $\xib$ goes negative. Finally, we use a 
synthetic NSR (whose form is derived in an Appendix) in section \ref{demo}, to 
demonstrate the above results.

\section{Mathematical preliminaries}
\label{sec:Math}

We will consider the evolution of a system starting from Gaussian 
initial conditions, with an initial power spectrum $P_{\rm{in}}(k)$. The 
two-point correlation function, $\xi(a,{\bf x})$, is defined as the Fourier 
transform of the power spectrum :
\beq
\xi(a,{\bf x}) = \int \frac{d^3{\bf k}}{\lb 2\pi\rb^3}P\lb a, {\bf k}\rb {\rm e}^{i {\bf k.x}} \;\; ,
\eeq
\noindent where $a$ is the expansion factor of the Universe ($a \propto t^{2/3}$ 
for an $\Omega = 1$ matter-dominated Universe). It is usually more convenient to work with 
the correlation function averaged over a sphere of radius $x$, given by
\beq 
\xib(a,x) = \frac{3}{x^3} \int_0^x \xi(a,y) y^2 dy \; .
\eeq
\noindent The power spectrum and the mean correlation function 
are related by the equations: 
\beq
\label{ptoxi}
\xib(a,x) = \frac{3}{2\pi^2 x^3} \int_0^\infty \frac{dk}{k} P(a,k) \left[ 
{\rm{sin}}(kx) - kx\;{\rm{cos}}(kx)\right]
\eeq
\noindent and 
\beq 
\label{xitop}
P(a,k) = \frac{4\pi}{3k} \int_0^\infty dx x \xib(a,x) \left[
 {\rm{sin}}(kx) - kx\;{\rm{cos}}(kx)\right] \; .
\eeq
\noi In the linear regime, it can be shown that $\xib \propto a^2$ so that 
$\xib_L(a,x) \propto a^2\xib_{\rm{in}}(a_i,x)$.\\
\noindent Note that the average two-point correlation function for
a power spectrum given by $P(k) = \delta_D(k - k_o)$, where $\delta_D$
represents the Dirac delta function is 
\beq
\xib(l) = \frac{3}{2\pi^2 l^3 k_o } \left[ {\rm sin}(k_o l) -
k_o l\;{\rm cos}(k_o l)\right] \; \; ,
\eeq
\noi which is negative over several ranges of scales, crossing zero for the 
first time when ${\rm tan}\lb k_ol \rb = k_ol$, {\it i.e.} $l \sim 4.493/k_o$. In 
fact, negative correlation functions always arise for sufficiently peaked 
functional forms of the power spectrum.

The two-point correlation function satisfies the equation of 
conservation of pairs \cite{LSS}:
\beq
\label{paircon}
\frac{\partial\xi}{\partial t} + \frac{1}{ax^2} \frac{\partial}{\partial x}
\left[ x^2\left( 1 + \xi \right) v \right] = 0 \; \; ,
\eeq
\noindent where $v(a,x)$ denotes the mean relative velocity of pairs at separation
$x$ and epoch $a$. Equation (\ref{paircon}) can be rewritten in terms of the mean 
2-point function, $\xib$, by defining a new dimensionless pair velocity, $h(a,x) 
\equiv -v/{\dot a}x$; this yields \cite{rajpad}
\beq
\label{Deqn}
\frac{\partial D}{\partial A} - h(A,X)\frac{\partial D}{\partial X} = 3h(A,X)
\eeq
where we have introduced the variables $D = {\rm{ln}}\;(1 + \xib)$, $A = {\rm{ln}}\;a$ 
 and $X = {\rm{ln}}\;x$.\\
\noi We will now assume $h \equiv h\left[\xib(a,x)\right]$; {\it i.e.} $h$ depends 
on $a$ and $x$ only through $\xib$. This is a standard assumption in the current 
literature \cite{ham,rajpad,mjw,pad96,padeng} and appears reasonably 
validated by numerical simulations. Our analysis and results depend  on this 
assumption, which may, of course, 
be looked upon as another way of truncating the BBGKY hierarchy. In the present 
work, this assumption will be treated as a basic postulate, validated by 
simulations; we will not address the question of its limits of validity.

Given the above assumption, $h \equiv h\left[\xib(a,x)\right]$, one can 
integrate equation (\ref{Deqn}) to obtain its general solution \cite{rajpad}, 
subject to the condition that this reduce to the form $\xib \propto a^2$ for 
$\xib \ll 1$. The characteristics of  equation (\ref{Deqn}) satisfy the constraint 
\beq 
\label{lxeqn}
x^3\left[1 + \xib(a,x)\right] = l^3 \; \; ,
\eeq
where $l$ is some other length scale. In the linear regime, $\xib \ll 1$ and $l 
\approx x$. At later stages of the evolution, however, as $\xib$ increases, equation 
(\ref{lxeqn}) shows that the scale $x$ becomes smaller and smaller as compared 
to the scale $l$. Thus, the evolution of clustering at a scale $x$ is determined 
by the original linear power spectrum at the (larger) scale $l$; this suggests 
that the true non-linear correlation function $\xib(a,x)$ at some scale $x$ 
can be expressed in terms of the linear correlation function $\xib_L(a,l)$, 
evaluated at a different scale $l$. The general solution is expressible as a 
non-linear scaling relation between $\xib(a,x)$ and $\xib_L(a,x)$, with 
$l$ and $x$ related by equation (\ref{lxeqn}); this can be written as \cite{rajpad}
\beq
\label{xilxi}
\xib_L(a,l) = {\rm{exp}} \left[ \frac{2}{3}\int^{\xib(a,x)} 
\frac{dz}{h(z)(1 + z)}\right] \;\; ,
\eeq
with $l = x[1 + \xib(a,x)]^{1/3}$. \\
\noi Given the form of $h(\xib)$, one can now relate the non-linear correlation 
function to the linear one; this form is usually obtained from simulations. 
However, it can also be shown from general theoretical arguments \cite{pad96} 
that $\xib(a,x)$ can be expressed in terms of $\xib_L(a,l)$ by the relations 
\setlength{\mathindent}{1.6 cm}
\begin{displaymath}
\xib_L(a,l) \:\: \:\:\:\:\:\:\:\:\:\;(\xib_L,\xib \ll 1)
\end{displaymath}
\setlength{\mathindent}{0.0 cm}
\beq
\xib(a,x) \:\: \propto \:\: \xib_L(a,l)^3 \;\:\:\:\:\:\:\:\: (1 < \xib < 200, 1 < \xib_L < 5.85)
\eeq
\setlength{\mathindent}{1.6 cm}
\begin{displaymath}
\xib_L(a,l)^{3/2} \:\: \:\:\:\:(200 < \xib, 5.85 < \xib_L) \; \; ,
\end{displaymath}
\setlength{\mathindent}{0.0 cm}
\noi if we confine ourselves to models with $\xib_L > 0$ everywhere. More 
exact fitting functions, valid over the entire range of $\xib > 0$, can be 
obtained from simulations \cite{ham,peacock94,mjw}, for example, the functional 
form given by Hamilton et al. (1991):
\beq 
\label{NSR}
\xib(a,x) = \frac{z + 0.358z^3 + 0.0236z^6}{1 + 0.0134z^3 + 0.0020z^{9/2}} \; \; ,
\eeq
where $z = \xib_L(a,l) \equiv a^2 \xib_{\rm in}$, and $\xib_{\rm in}$ is the 
initial mean correlation function. \\
\noi Thus, given an initial power spectrum $P_{\rm{in}}(k)$, one can determine 
the average correlation function (and, hence, the power spectrum, using equation 
(\ref{xitop})) at any epoch, by simply using equations (\ref{ptoxi}), (\ref{lxeqn}) 
and the non-linear scaling relation (\ref{NSR}).

There is, however, one serious problem with these fitting functions.
The original analysis, leading to equations (8) and (9) made no assumptions
regarding the sign of $\xib_L$ and is valid even if $\xib_L < 0$ at some
scales. The fitting functions, however, {\it require} that the linear mean correlation
function be everywhere positive. This is partly due to the fact that standard
power spectra used in cosmology (such as the CDM power spectrum and its variants)
do have positive definite correlation functions and there appeared to be no 
compulsion to consider the case of negative $\xib$. We feel, however, that 
this situation is unsatisfactory for two reasons : \\
        (i) The NSR has proved to be a valuable tool in understanding the
physics of gravitational clustering and --- in particular --- the transfer
of power between two scales. The basic question in the study of transfer of power
is how a sharply peaked power spectrum evolves due to gravitational clustering.
To answer this question using the NSR, we need to generalise it for negative
values of the mean correlation function.\\
        (ii) As we will show in section \ref{obs}, observations cannot
rule out the existence of negative $\xib$ over certain ranges of scales 
in the universe. Since this feature has very important implications, 
it should be taken seriously, without any theoretical prejudice.

It should be noted that attempts have also been made to write NSRs directly for the
function $\Delta^2(k) \equiv k^3 P(k)/(2\pi^2)$ in frequency space
\cite{peacock94,peacock}, rather than in terms of the mean correlation function
$\xib$, as discussed above. In such Fourier space NSRs, one writes
\beq
\label{eqn:kNSR1}
\Delta^2_{NL}(k_{NL}) = F[\Delta^2_{L}(k_{L})] \ ,
\eeq
\noi with the wavenumbers $k_{NL}$ and $k_{L}$ related by
\beq
\label{eqn:kNSR2}
k_{L} = [1+\Delta^2_{NL}(k_{NL})]^{-1/3} k_{NL} \ .
\eeq
For example, Peacock \& Dodds (1994) give the following form for the function
$F$ (we have put $g(\Omega) = 1$ in their expression, for the case of an
$\Omega = 1$, matter-dominated Universe) :
\beq
\label{eqn:kNSR3}
F(x) =x \; \left[{ 1+0.2\beta x +(A x)^{\alpha\beta} \over
1 + ([A x]^\alpha /[11.68 x^{1/2}])^\beta}\right]^{1/\beta} \ ,
\eeq
with $A = 0.84 $ and $\alpha = \beta=2$. Since $\Delta^2(k) \ge 0$, $k_{NL} \ge k_{L}$,
and power is transferred from small to large wavenumbers, i. e. from large to small
scales. However, no mathematical basis presently exists for scaling relations for 
quantities other than the mean correlation function $\xib$; this is discussed in 
more detail in the next section. 

\section{NSRs for different statistical indicators.}
\label{nsr_compare}

We discuss, in this section, a few important general issues related to the
role of $\bar\xi$ in the context of non-linear scaling relations (NSRs). 
Throughout this paper we will use the term NSR to mean a 
relationship between the exact value of some statistical indicator $Q$ 
(which could be $\bar\xi, \xi, \Delta \equiv k^3P(k)/2\pi^2$, etc) to its value 
$Q_{lin}$ as given by linear theory, in the form $Q(x)=F[Q_{lin}(l)]$ with 
$l=(1+Q)^{1/3}x$, where $x$ and $l$ are two scales and $F$ is a prescribed 
function. In the case of Fourier space quantities like $\Delta(k)$ or $P(k)$, 
we will, of course, interpret $x$ and $l$ in term of $k^{-1}$ for the two scales.

There are two broad perspectives one can take regarding such NSRs. In the first 
approach, NSRs can be treated as a set of relations which are of 
considerable practical utility in studying the evolution of clustering.
In this approach, they are merely convenient sets of approximate rules by
which non-linear quantities can be obtained from linear ones, thereby 
facilitating a comparison between theory and observations. Taken in this 
spirit, the key issues in this area are only the accuracy of the fitting 
functions $F$ for the NSR, the dependences on various parameters and, finally,
 given an NSR for a particular quantity $Q$, how best to write down (purely 
as fitting functions) NSRs for other statistical indicators. There are no 
fundamental issues; NSRs exist entirely by accident, but are of use in 
studying non-linear structure formation.

The above attitude, however, is tantamount to sweeping the entire issue under 
a carpet, without much investigation. In fact, it appears 
that NSRs are indicative of a key (and not completely understood) feature 
of non-linear gravitational clustering and hence need to be investigated 
thoroughly. In support of this view, we would like to stress the following :

(i) Though the original NSR for $\bar\xi$ was obtained as a fitting function 
to simulation data by Hamilton et al. (1991), the work by Nityananda and Padmanabhan 
(1995, hereafter NP) clearly spelled out its theoretical origins. In particular, 
this work showed that (a) there exists an exact equation satisfied by 
$\bar\xi$ whose integral curves lead to the relation $l^3=x^3(1+\bar\xi)$ 
(b) the only key assumption which is needed to obtain the NSR for 
$\bar\xi$ is that $h$ is a function of $\bar\xi$ alone. 

(ii) So far, there has been no evidence for any mathematical foundation for 
NSRs for other statistical indicators like $\Delta^2$ or even $\xi$. In fact, 
there is no analogue for local differential equations like equation (25) in 
NP, for other statistical indicators like $\Delta^2$ or the power spectrum. 
Thus, NSRs appear to exist {\it only} for the mean correlation function 
$\bar\xi$, implying that $\bar\xi$ enjoys a special status as an indicator of 
clustering. The NSRs given in the literature for $\Delta(k)$, for example,
arose purely as an afterthought, guided by the nature of the NSR for $\bar\xi$.

(iii) Given the form of $\bar\xi$, it is possible to obtain $\Delta^2$ and $\xi$
by simple analytic procedures. However, {\it if} the NSR is true for $\bar\xi$, it 
can be shown that such a similar NSR --- as defined above --- {\it cannot} be 
exactly satisfied for $\Delta^2$ or $\xi$. While there have been attempts in the 
literature to obtain fitting functions for $\Delta^2_{NL}$ in terms of $\Delta^2_L$ 
(Peacock \& Dodds 1994, 1996), along the lines of the NSR for $\bar\xi$, such 
exercises do not have the same level of fundamental validity as the NSR for the 
mean correlation function.

The above distinction is not usually of great relevance for the following reasons :
First, if the power spectrum is a smooth, mildly varying, power law with
adequate asymptotic properties, an NSR for $\bar\xi$ will lead to similar
approximate relations for $\Delta^2$, $\xi$, etc. Which of them is used in a 
specific context --- assuming that this is all one uses the NSR for --- is purely 
a matter of convenience. Further, since any fitting function is only approximate, 
the fact that some NSRs are more approximate than others is often not relevant 
in practical situations.

The situation, however, is very different when one moves away from simple-minded 
power spectra and considers a more generic situation. The original analysis 
of NP (which provided the theoretical foundation for NSRs) did not make any 
assumptions about the nature of the initial mean correlation function or the 
power spectrum. Consider, for example, the case of a power spectrum which is 
sharply peaked around some value, such as a Dirac delta function or a Gaussian. 
For such a power spectrum, $\Delta^2$, $\xi$ and $\bar\xi$ {\it have very different 
shapes}. In particular, $\bar\xi$ is quite flat at small scales and decreases 
towards large scales, while $\Delta^2$ is close to zero at small scales. Since 
the shapes of $\bar\xi$ and $\Delta^2$ are different, the evolution predicted 
by using the NSR in real space and Fourier space will be totally different. 
This can be seen very clearly in the case of a power spectrum which is a Dirac 
delta function or for any power spectrum with a cutoff at short wavelengths (i.e. 
$P(k)=0$ for $k > k_c$, with some finite $k_c$). It is precisely in such situations 
that the mathematical foundations for the two NSRs come into question and, given 
the fact that the NSR for $\bar\xi$ has a theoretical basis while that for 
$\Delta^2$ is only a fitting function, it is the former which must be used when 
the results from the two differ. Numerical simulations of peaked power 
spectra (Bagla \& Padmanabhan 1997) have shown that, even in such cases, power 
is indeed transferred from larger to smaller scales, as indicated by the NSR 
in real space. In fact, such power transfer to scales smaller than the cutoff in 
the power spectrum {\it would be impossible} if the Fourier space NSR were 
correct, simply from the fact that the power spectrum at a given value of $k$ is 
a continuous function of the scale factor (this result is true for any power 
spectrum with a short wavelength cutoff, not merely a peaked power spectrum). 
While the power spectra of Bagla \& Padmanabhan (1997) were exponentially damped 
beyond a small range of scales, and hence did not have a sharp cutoff, the fact 
that these simulations did show the transfer of power to small scales, as expected 
from the $x-$space NSR, also indicates that the correct NSR to use is the one for 
real space, {\it not} Fourier space. 

We note finally, that even if theoretical foundations indeed existed for both forms of 
the NSR and the two yet yielded different results for the same initial power 
spectrum / correlation function, one should take resort to dynamical tests such as 
N-body simulations to ascertain which of the two is applicable. In fact, such a 
situation might well prove interesting as it would provide information about the 
dynamical correctness of the two scaling relations.

The strength of the present work is that its conclusions do not depend on the form 
of the NSR but arise entirely from its existence. The synthetic NSR of the last section 
of the paper (and the appendix) is used solely to illustrate the results. It should 
be emphasised that we will show analytically that the results are independent of the 
form of the NSR. In the spirit of the above discussion, we will work in $x$-space and 
not in Fourier space during the analysis.

\section{An aside : Observational constraints}
\label{obs}

As mentioned earlier, negative values of $\xib$ are {\it not} 
mathematically forbidden and, in fact, arise naturally for sufficiently 
peaked forms of the power spectrum. We, will, in this section, briefly review
observational constraints to investigate whether present-day observations 
of the correlation function or the power spectrum rule out the existence of 
negative $\xib$ in our universe. 

At very large scales, $l \ga 1000 h^{-1}$ Mpc, 
the CMBR anisotropy measurements provide a constraint on the slope and 
amplitude of the power spectrum; $\xib_L$ is probably positive in this range. 
At small scales, $1 h^{-1}$ Mpc $ \la l \la  200 h^{-1}$ Mpc, the galaxy-galaxy 
correlation function provides the shape and amplitude of the baryonic component of 
the universe. This can be converted into providing some handle on the underlying 
dark matter distribution and the results are consistent with a positive mean 
correlation function. At still smaller scales, information on the power 
spectrum arises from the study of abundances of bound structures (like quasars, 
damped Lyman-$\alpha$ systems, etc) and it is very likely that the mean correlation 
function is again positive at these scales. There is, however, a range of scales 
($ \sim  200 h^{-1}$ Mpc -- $1000 h^{-1}$ Mpc) which are not directly probed by 
present observations 
and one can easily construct power spectra for which the correlation function 
goes negative in this range. Of course, while doing this, one has to also ensure 
that the additional power does not create any observable consequences at other 
scales. 

We demonstrate, next, (by an explicit example) that it is possible to construct 
power spectra which satisfy the above observational constraints and yet yield 
negative $\xib$. As mentioned above, the observations allow negative values of 
$\xib$ over the range $200 h^{-1}$ Mpc $< x < 1000 h^{-1}$ Mpc. Consider the 
COBE-normalised CDM power spectrum of Bardeen et al. (1986) (eq. G3) 
\beq
P_{\rm{CDM}}(k) = Ak \left[ \frac{{\rm {ln}}\left[1 + 2.34q\right]}{2.34q} \right]^2 
\;\;\times
\eeq
\begin{displaymath}
\; \; \; \; \; \; \; \; \; \; \; \; \; \; \;\left[1+3.89q + (16.1q)^2 + (5.46q)^3 + (6.71q)^4\right]^{-1/2} \; \; ,
\end{displaymath}
\noi with $q = k/(\Omega h^{2}{\rm Mpc}^{-1})$ (using $\theta = 1$ in eq. (G3), 
corresponding to photons and three types of relativistic neutrinos). The value of 
$A$ is set by COBE normalisation to be about ($24 h^{-1}$ Mpc)$^4$. A good fit 
to the average correlation function $\xib_{\rm{CDM}}$ for the above spectrum is 
given by Hamilton et al. (1991) \\
\beq
\xib_{\rm {CDM}}(x) = 0.51 \left[ {\rm {ln}}\left( 1 + \frac{5}{x} \right)\right]^3 
\frac{{\rm ln}\left(1 + x/12\right)}{x/12} \; \;\times 
\eeq
\begin{displaymath}
\: \: \: \: \; \; \; \; \; \; \; \; \; \; \; \; \; \; \; \; \;\frac{ 1 + 0.394x + 0.00316x^2} {1 + 0.142x + 0.00129x^2} \;\; ,
\end{displaymath}
\noi where $x$ is in $\Omega^{-1}h^{-2}$ Mpc and $h$ is the Hubble parameter. 

We will work in an $\Omega = 1$ Universe and add to the CDM spectrum, a 
peaked spectrum of the form 
\beq 
P_2(k) = A' k^4 e^{-\lambda k} \; \; ;
\eeq
the correlation function corresponding to $P_2(k)$ is 
\beq
\xib_2(x) = -\frac{72 A' \lambda}{\pi^2} \frac{3x^2 - 5\lambda^2}
{\left( x^2 + \lambda^2\right)^5} \;\; ,
\eeq
which clearly goes negative for $x > (5/3)^2 \lambda$. 

\begin{figure}
\centering
\psfig{file=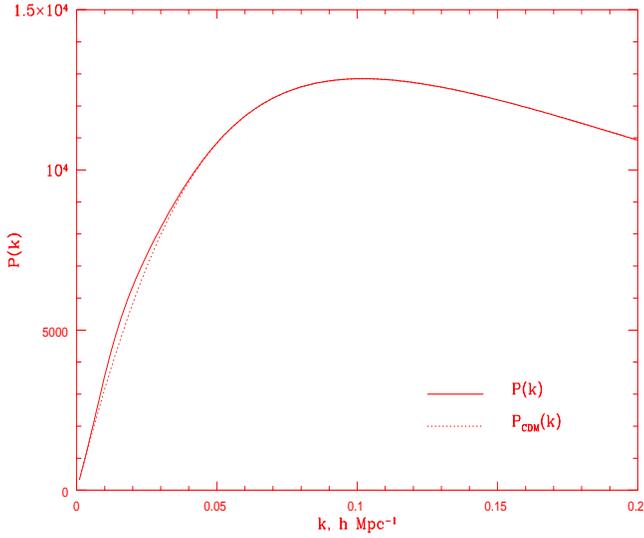,width=3.5truein,height=3.0truein,angle=-90}
\caption{The figure shows the CDM power spectrum (dotted line) and the net 
power spectrum $P(k) = P_{\rm {CDM}}(k) + P_2(k)$ (solid line). It can be 
seen that the two are different only over a small range in $k$-space.}
\label{fig7}
\end{figure}

The net power spectrum $P(k) = P_{\rm {CDM}}(k) + P_2(k)$ (with $A' \sim
\lb 841 h^{-1} \;{\rm Mpc}\rb^4$ and $\lambda = 250\; h {\rm Mpc}^{-1}$) is 
plotted in figure (\ref{fig7}) (solid line), along with the CDM spectrum 
(dotted line). It can be seen that the deviation of the net spectrum from 
the CDM spectrum occurs only over an extremely small range of $k$ values and this 
deviation is itself extremely small. 

\begin{figure}
\centering
\psfig{file=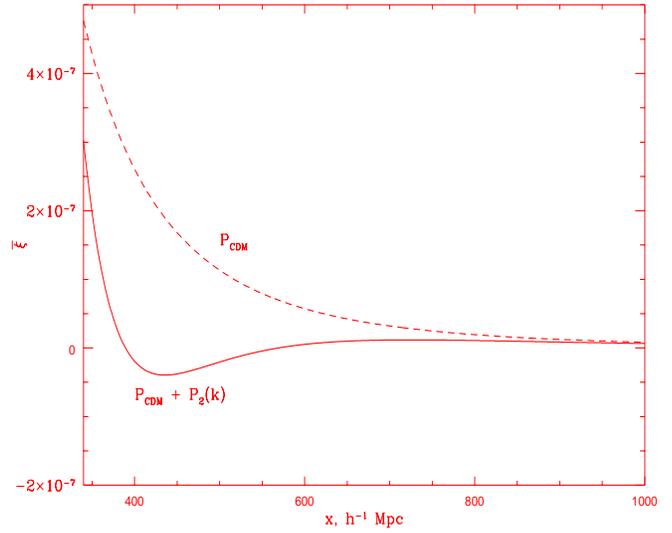,width=3.5truein,height=3.0truein,angle=-90}
\caption{The average correlation function $\xib(x)$, for the two power 
spectra of figure (1). The correlation function corresponding to the total 
power spectrum, $P(k)$, can be seen to go negative at $x \sim 390 h^{-1}$ Mpc. 
The two correlation functions approach each other beyond $ x \sim 800 h^{-1}$ Mpc.}
\label{fig8}
\end{figure}

Figure (\ref{fig8}) shows the average correlation 
function $\xib(x)$, corresponding to $P(k)$. It can be seen that $\xib(x)$ becomes 
negative at $x \sim 390 h^{-1}$ Mpc and then returns to positive values at $x \sim 
 600 h^{-1}$ Mpc. In the region probed by COBE ($1000 h^{-1}$ Mpc -- $3000 h^{-1}$ Mpc), 
$\xib$ is indistinguishable from the CDM correlation function; this is due to the 
extremely rapid fall-off in $\xib_2(x)$. 

Thus, even broad band power spectra can yield negative correlation functions. 
The example shown in figure (\ref{fig7}) is just one of many possible ways 
to construct such power spectra, which agree with the current observations. It is 
hence clearly important to understand the effects of such anti-correlated regions 
on the evolution.\\
We emphasise that the power spectra we have constructed will satisfy all 
observational constraints currently available but will introduce some new
features in the evolution of the universe. The purpose of doing this is only to
stress how little we know about our universe; we have hence not worried
about whether there exist suitable physical processes which will lead to 
such a power spectrum. 

\section{Continuity arguments and bound on $\xib$}
\label{negxi}

We will now consider the evolution of $\xib$ with time, at a given scale $x$, 
and show that the mere existence of a negative value of $\xib_L$, at some scale $l$, 
{\it implies} an upper limit to the value of $\xib$ which can be reached at
any scale smaller than $l$; further, the maximum value $\xib_{\rm{max}}$ at 
any scale is a function of the scale itself and the location $L_0$, where 
$\xib_L$ first crosses zero. This effect also results in the correlation function 
acquiring a $x^{-3}$ dependence at late stages of the evolution.

We begin by noting that the arguments leading to the paradigm for transfer 
of power (see Section 2) made no assumption about the sign of $\xib$; 
these arguments, as well as equations (\ref{lxeqn}) and (\ref{xilxi}), thus 
hold good for the negative regime as well. Next, the relation 
\beq
\label{lxeqn2}
l = x\left[ 1 + \xib (a,x)\right]^{1/3} \; \; ,
\eeq
\noi which governs the tranfer of power from one scale to another, implies that a given 
value of $\xib$ can arise at some scale $x$ {\it only} by the transfer of power 
from a single scale $l$. For $\xib > 0$, $x < l$, {\it i.e.} power flows to smaller 
scales, while, if $\xib < 0$, $x > l$, {\it i.e.} power transfers to larger scales. 
We stress that equation (\ref{lxeqn2}) made no assumptions about the sign of 
$\bar\xi$ except, of course, the obvious condition, $(1+\bar \xi)>0$.

Let $L_0$ be the first scale at which $\xib_L$ crosses zero and let $x_0$ be the 
scale of interest, with $x_0 < L_0$. Initially, in the linear regime, $\xib \ll 1$ , 
$x_0$ is influenced by scales $l$ such that $l \approx x_0$. At later epochs, 
as $\xib$ increases, $x_0$ is influenced by larger and larger scales, upto the scale 
$L_0$. Power from the scale (infinitesimally larger than) $L_0$, however, must be 
transferred to scales {\it larger} than $L_0$, since $\xib_L < 0$ here; clearly, 
this scale $(l = L_0 + \Delta l)$ {\it cannot} influence $x_0$. However, as has 
already been emphasised, a given value of $\xib$ can arise at the scale $x_0$ only 
by the transfer of power from a single scale; this immediately implies that $\xib(x_0)$ 
can {\it never} exceed the value $\xib_o$, given by
\beq
\xib_o = \left[ \frac{L_0}{x_0} \right]^3 - 1 \; \; ,
\eeq
since this would require power transfer from the scale $L_0 + \Delta l$ to $x_0$, 
which is impossible. Similarly, if $\xib_L < 0$ over some range of 
$l$, say, from $l_1$ to $l_2$, then no value of $\xib$ which satisfies the relation 
\beq
\xib = \left[ \frac{l'}{x_0} \right]^3 - 1
\eeq
\noi can be reached at $x_0$, for $l'$ between $l_1$ and $l_2$.\\
\noi The correlation function, however, is a continuous function; clearly, then, 
if some values of $\xib$ cannot be reached, $\xib$ {\it must} be bounded from above 
by the lowest of these values. Thus, the maximum value which $\xib$ can take at some 
scale $x$ ($x < L_0$) is given by 
\beq
\label{ximaxeqn}
\xib = \left[ \frac{L_0}{x} \right]^3 - 1 
\eeq
\noi where $L_0$ is the smallest scale at which $\xib_L$ crosses zero.
Thus, at late times, $\xib \gg 1$, we must have 
\beq
\xib \propto x^{-3}
\eeq
\noi for $x$ smaller than $L_0$. This is a generic result, independent of the 
exact form of the initial correlation function, except for the requirement 
that it become negative at some scale. 

Note that $\xib(x) \propto x^{-3}$ over some range of scales implies that 
$\xi(x) = 0$ over this range (and vice-versa). For example, if $\xi(x) \approx 0$ 
for $x > L$, then 
\begin{eqnarray}
\xib(x > L) &=& \frac{3}{x^3}\left[ \int_0^L y^2 \xi(y)dy + \int_L^x y^2 \xi(y)dy\right]\\
&\simeq& \frac{3}{x^3}A \;\; ,
\end{eqnarray}
\noi where $A =  \int_0^L y^2 \xi(y)dy$ is a constant. Thus, $\xib(x) \propto x^{-3}$, 
for $x > L$, if $\xi(x) \approx 0$ here. The result that the relation $\xib(x) 
\propto x^{-3}$ spreads to smaller scales at late times can be then seen to imply 
that $\xi(x)$ becomes negligible at smaller and smaller scales at later and later epochs.

It may appear strange that the mere existence of negative $\xib_L$ has far-reaching 
effects on the evolution. However, it must be emphasised that the positive and 
negative $\xib$ regimes are {\it not} equivalent, due to the lower bound of $(-1)$ 
on $\xib$ in the negative regime; no such bound exists for $\xib > 0$. This implies 
an asymmetry in the very structure of the equations, if one uses $\xib$ as a 
variable. In fact, this also indicates that Log[$1 + \xib$] is probably the correct 
variable which should be used; in this case, the above asymmetry does not arise, 
since Log[$1 + \xib$] takes all values from $-\infty$ to $+\infty$. 

In general, linear theory ceases to be valid when $|\xib|$ is of the 
order of unity. However, while using the NSR, one maps the linearly 
extrapolated value $\xib_L$ to the actual value of $\xib$. This poses no 
mathematical problems for $\xib > 0$, since arbitrarily large positive values of 
linearly extrapolated $\xib_L$ are {\it mathematically} allowed and 
map to still higher values of actual $\xib$ through the NSR; the situation 
is quite different in the negative regime since the actual $\xib$ 
is constrained to be greater than $(-1)$. This asymmetry appears to play a role 
in situations in which negative correlations exist.

We mentioned in section \ref{nsr_compare} that power transfer to scales smaller
than the cutoff in the initial power spectrum is impossible for power spectra with sharp
cutoffs, if the Fourier space NSR is correct. This can be seen by an argument similar
to the one discussed above. Consider an initial power spectrum with a short wavelength
cutoff at some scale $k_o$. In Fourier space NSRs (see equations (\ref{eqn:kNSR1}) to
(\ref{eqn:kNSR3})), the power $P(k)$ at a wavenumber $k$
at some epoch originates at a fixed scale $k_l$, with $k_l$ and $k$ related by 
\cite{peacock94}
\beq
\label{eqn:k-kl}
k^3 = k_l^3 \left[ 1 + \Delta^2(k) \right] \ ,
\eeq

\noi with $\Delta^2(k) = k^3 P(k)/(2\pi^2)$. Since $\Delta^2(k) \ge 0$, clearly
$k_l \le k$. If we now consider a wavenumber $k$ beyond the cutoff in the initial
spectrum, with
$k > k_o$, there exists a range of wavenumbers between $k$ and $k_o$, where
$P_{in}(k) = 0$ and the power at $k$ {\it is then forbidden to take values which could
originate between} $k_o$ and $k$. Since the power spectrum is a continuous
function of the scale factor, if $P(k)$ is not allowed to take a set of values,
it also cannot take values larger than the lowest value of this set. Thus,
even in Fourier space, it can be seen that an upper bound exists on the power
spectrum at large wavenumbers (i.e. small scales) $k > k_o$, for initial power
spectra which cut off at some wavenumber $k_o$. We note that such power spectra
do, in fact, give rise to negative correlation functions; one could thus have
instead used the real space NSR to demonstrate the upper bound on $\xib$. However,
there are also other power spectra such as Gaussians, which do produce negative
$\xib$ even without sharp cutoffs in the initial spectrum; in such cases, the
upper bounds are not as obvious from a Fourier space analysis but can be clearly
seen in real space. Of course, as discussed in section \ref{nsr_compare},
different results may then be obtained for the late-time behaviour of $P(k)$ and $\xib$
from the two NSRs, especially in the case of peaked initial power spectra. In such
situations, $\xib$ is picked out as the statistical indicator to be used in the NSR,
by the analysis of NP which provides a theoretical basis for the real space scaling
relations.

\section{NSR for negative $\xib$}
\label{demo}

We require an NSR for the negative $\xib$ regime to explicitly illustrate the 
results of the preceding section. However, the form of the NSR given 
in equation (\ref{NSR}) tacitly assumes that $\bar\xi_L(a,l) >0$ for all 
$l$; as mentioned earlier, this requirement has no fundamental significance. 
The above form, however, can be derived from theoretical arguments 
related to spherical collapse, in the regime $\xib > 0$ \cite{pad96}. 
This suggests that the NSR for the case $\xib < 0$ may also be 
obtained by an analysis of a spherically symmetric model, with negative 
correlations at some scale. This will allow us to guess at an ansatz for 
$h(\xib)$, based on the relation between $h$ and the density 
contrast $\delta$ for this model. This form for $h(\xib)$ will then be used 
in equation (\ref{xilxi}) to obtain an NSR in the negative $\xib$ regime. 

We will use the ansatz, 
\beq
h(\xib) = \frac{2}{3} \xib + \frac{1}{6}\xib^2 \;\;\;\;\;\;\;\; (-1 < \xib < 0) \;\; ,
\eeq
\noi for $h(\xib)$ (see Appendix for motivation). Substituting for $h(\xib)$ in equation 
({\ref{xilxi}), we obtain, after some algebra
\beq
\xib_L(a,l) = \frac{\xib\left(1 + \xib/4\right)^{1/3}}{(1 + \xib)^{4/3}}\;.
\eeq
\noi The inverse mapping, giving $\xib$ in terms of $\xib_L$, can be fitted to
better than 5\% by the relation
\beq
\label{newNSR}
\xib = \frac{\xib_L + 0.88\xib_L^2 + 2.086\xib_L^3}{1 + 1.798\xib_L^2 - 2.086\xib_L^3} 
\;\;\;\;\;\;\; (-1 < \xib < 0) 
\eeq
\noi Equations (\ref{lxeqn}) and (\ref{newNSR}) implicitly determine $\xib(a,x)$
in terms of the linearly evolved correlation function, $\xib_L(a,l)$ for all
values, positive and negative, of $\bar \xi_L$. 
We mention that the above NSR, equation (\ref{newNSR}), will be used solely 
for the purpose of illustrating the results of Section (4), since these results 
are independent of the exact form of the NSR. The true mapping between $\xib$ 
and $\xib_L$ in this regime should, of course, be determined from N-body simulations. 

We will use the net power spectrum $P(k)$ of figure ({\ref{fig7}) (which is 
allowed observationally) to demonstrate the effects of negative $\xib$. The results are 
shown in figures (\ref{fig3}) and (\ref{fig4}), in which $\xib$ is plotted against 
the scale factor, for two scales, $x_1 = 150 h^{-1}$ Mpc and $x_2 = 20 h^{-1}$ Mpc. It can be seen 
that, in both cases, $\xib$ does not continue to increase with $a$ (or, equivalently, 
with time), but instead flattens out at late times and approaches a value of 
$(L_0/x_0)^3 - 1$ (dotted lines), as predicted by the analysis. Here, $L_0$ is the
scale at which $\xib_L$ first goes negative and $x_0$ is the scale under 
consideration. We emphasise that the mapping for negative $\xib$ has no
influence on the results; any mapping for this regime would have yielded the same
results since these stem entirely from the {\it existence} of negative values of
the correlation function. 

Figure (\ref{fig5}) shows a log-log plot of $\xib$ versus $x$
at four different epochs; the solid line has a slope of $-3$. It can be seen that 
more and more scales acquire the $\xib \propto x^{-3}$ behaviour with time. Scales 
with $x \la 3 h^{-1}$ Mpc ({\it i.e.} $\xib \ga 2 \times 10^6$) have not as yet (upto $a = 
5 \times 10^5$) been influenced by the region with $\xib_L < 0$ and will hence only 
acquire the $x^{-3}$ slope at later epochs. It should be emphasised that the above 
discussion regarding the upper bound on $\xib(x)$ is applicable {\it only} for scales 
$x$ which are {\it smaller} than the scale $L_0$ at which the initial mean correlation 
function $\xib_L$ first goes negative and not for scales $x > L_0$. We do not discuss 
the evolution of $\xib(x)$ for $x > L_0$ here, since this would depend on the exact 
form of the scaling relation in the negative $\xib$ regime and we presently only 
possess a rather synthetic NSR in this region.

\begin{figure}
\centering
\psfig{file=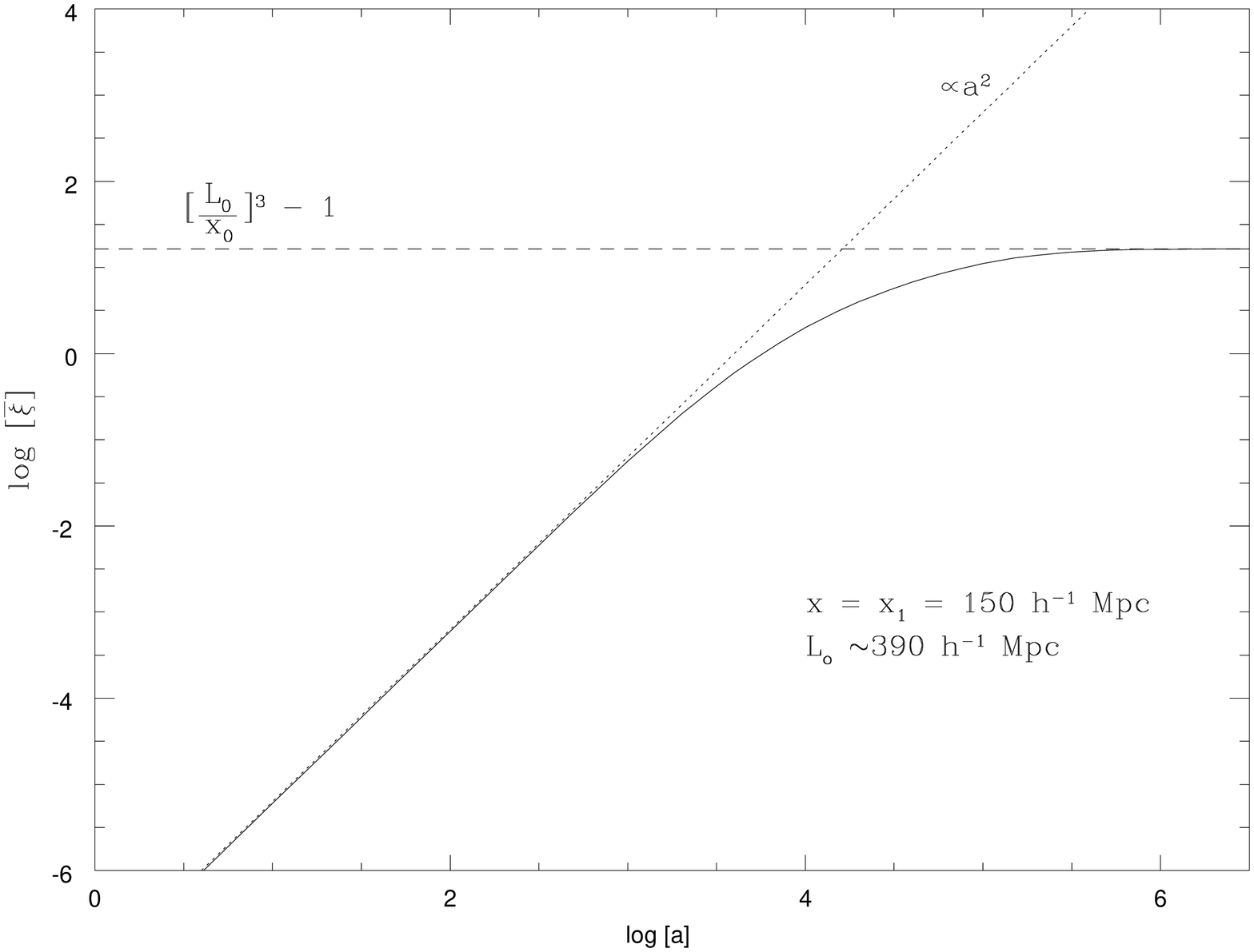,width=3.5truein,height=3.0truein,angle=-90}
\caption{The figure shows $\xib$ as a function of scale factor, at $x_1 = 150 h^{-1}$ Mpc, 
for the power spectrum $P(k)$ discussed in the text. The dotted line corresponds to 
$\xib \propto a^{-2}$.} 
\label{fig3}
\end{figure}

\begin{figure}
\centering
\psfig{file=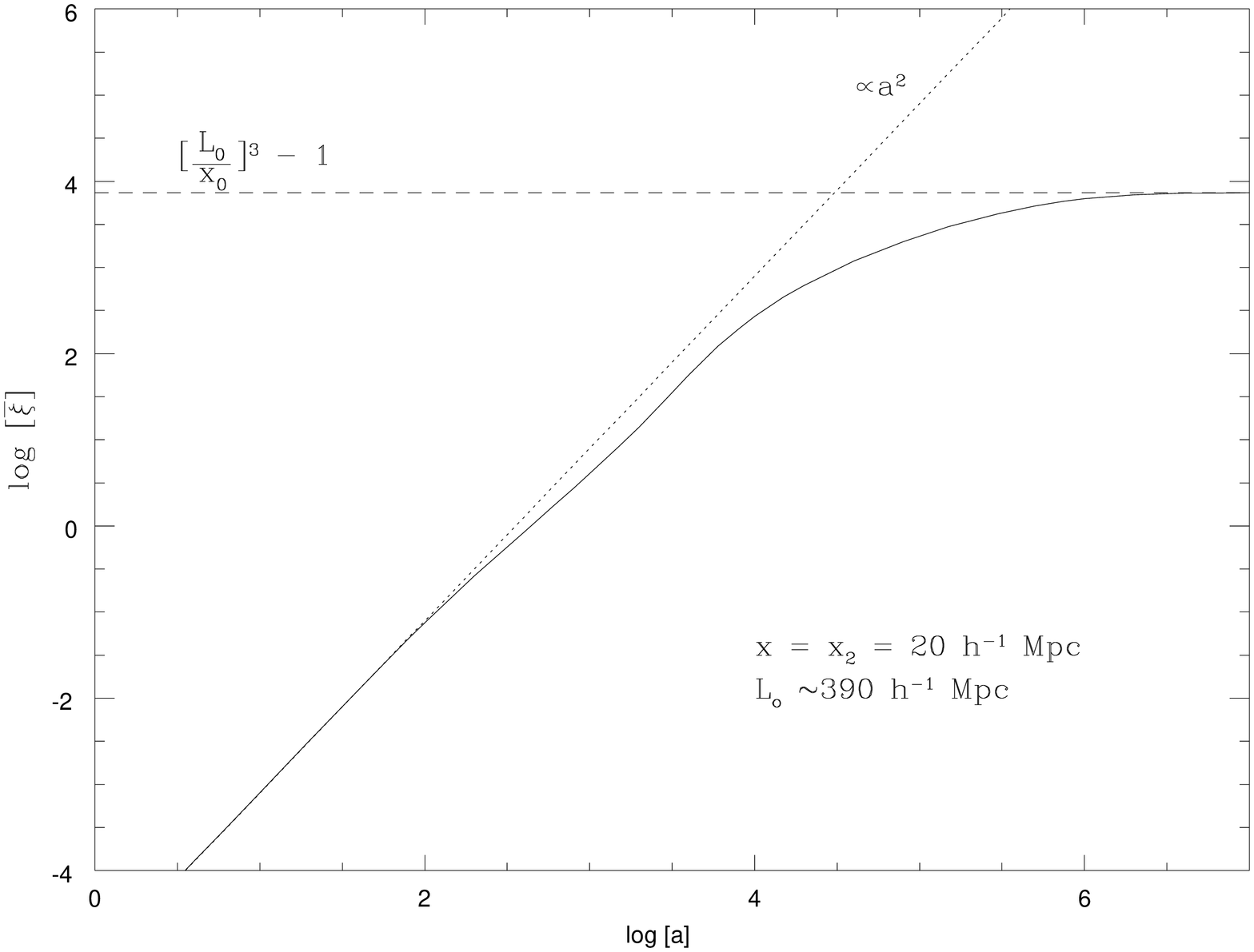,width=3.5truein,height=3.0truein,angle=-90}
\caption{The figure shows $\xib$ as a function of scale factor, at $x_2 = 20 h^{-1}$ Mpc, 
for the power spectrum $P(k)$ discussed in the text. The dotted line corresponds to 
$\xib \propto a^{-2}$.}
\label{fig4}
\end{figure}

\begin{figure}
\centering
\psfig{file=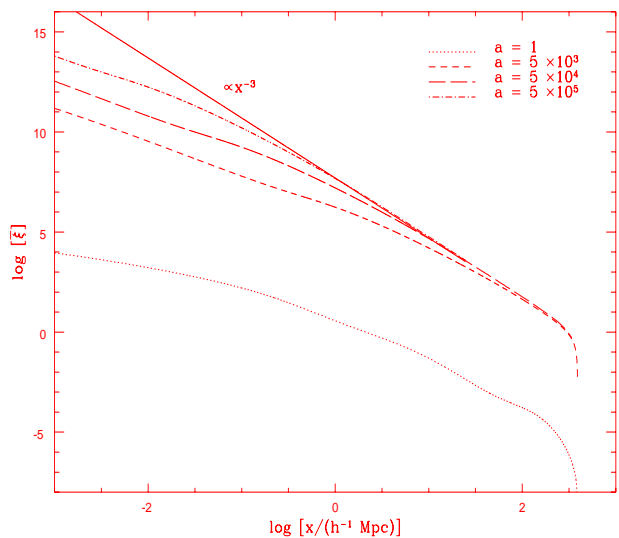,width=3.5truein,height=3.0truein,angle=-90}
\caption{The figure shows the mean correlation function $\xib$ plotted
against scale $x$, for four different epochs (dot-dashed, dotted, short-dashed 
and long-dashed lines). The solid line has a slope of $-3$. It can be seen that 
more and more scales acquire the $\xib \propto x^{-3}$ behaviour at later and 
later times.}
\label{fig5}
\end{figure}

Finally, we note that the ``asymptotic'' behaviour of $\xib$ does not arise
at any particular value of $\xib$ but depends only on the scales in question, $x_0$ 
and $L_0$. For example, in figure (\ref{fig3}), the asymptotic behaviour arises when $\xib
\sim 16$ ({\it i.e.} in the quasi-linear regime) while, in figure (\ref{fig4}),
this behaviour arises when $\xib \sim 7 \times 10^3$ ({\it i.e.} in the non-linear regime).

The mean correlation function $\xib$ is known to scale as $\xib \propto a^{2}$, 
$a^{6}$ and $a^{3}$ (in 3 dimensions) in the linear, quasi-linear and non-linear 
regimes of structure formation \cite{ham,pad96}. However, the existence of 
negative correlation functions appears to give rise to a {\it fourth regime} in 
gravitational clustering, besides the above three regimes. Here, $\xib$ asymptotically 
approaches a constant value, at any scale shortward of the scale at which $\xib_L$ 
crosses zero. Clearly, high dynamic range simulations will be needed to test 
this result; this may well prove difficult, or even prohibitive, in 3-dimensions. 
Such high dynamic ranges can, however, be attained in 2-D simulations; 
we note that earlier 2-D simulations \cite{bep} have shown that $\xib$ does not 
continue to grow as $a^{2}$ in the non-linear regime (as expected from stable 
clustering), but begins to flatten out at late times. It is possible that this is 
indicative of the existence of a fourth regime, beyond the stable clustering phase.  
{\tt Of course, the present results are quite independent of the stable clustering 
ansatz, since they stem solely from the existence of the non-linear scaling relation. }

We note finally that the evolution of non-linear structure may also be 
studied using other models; for example, one may view the non-linear 
correlation function as arising from the superposition of the density 
profiles of non-linear halos \cite{seljak,peacock2000}. However, 
NSRs seem to be reasonably validated by numerical simulations; we have 
hence not considered other models, such as the halo model, here, but have limited 
ourselves to attempting to understand the scaling relations themselves.

In summary, the mere existence of negative correlation functions have 
far-reaching effects on the evolution of clustering. They result in an 
upper bound on $\xib$ at scales smaller than the scale at which $\xib_L$
goes negative and also cause the relation $\xib \propto x^{-3}$ at these scales, 
at late times. Negative correlation functions are not forbidden observationally; 
it is hence important to use N-body simulations and/or semi-analytic arguments 
to obtain a form for the NSR in this regime. 

{\bf Acknowledgments }\\
It is a pleasure to thank K. Subramanian and R. Nityananda  for stimulating 
discussions during the course of the present work, as well as for a critical 
reading of an earlier draft of the paper. The research work of one of the 
authors (TP) was partly supported by the INDO-FRENCH Centre 
for Promotion of Advanced Research under grant contract No 1710-1. 
We also thank an anonymous referee for his comments on the paper.\\

\section{Appendix}
We will use the evolution of a spherically symmetric, compensated void 
to obtain some insight into the form of $h(\xib)$. Consider an initial density 
perturbation of the form 
\setlength{\mathindent}{1.2 cm}
\begin{displaymath}
\; \:\:\:\: -|\delta_1| \:\:\:\:\:\:\: (0\; <\; R\; <\; R_1)
\end{displaymath}
\setlength{\mathindent}{0.1 cm}
\beq
\delta(R) \:\:\:\:\:\: = \:\:\:\: \delta_2 \:\:\:\:\:\: (R_1\; <\; R\; <\; R_2)
\eeq
\setlength{\mathindent}{1.2 cm}
\begin{displaymath}
\;\;\; \:\:\:\;\; 0 \:\:\:\:\:\:\;\;\; (R_2\; <\; R) \; \; ,
\end{displaymath}
\setlength{\mathindent}{0.0 cm}
\noi such that the effects of the under-dense region within $R_1$ 
are ``compensated'' by the over-dense region between $R_1$ and $R_2$.
This implies that a shell with radius larger than $R_2$ does not feel any excess 
gravitational force, which would cause it to deviate from the smooth 
background expansion. Such a configuration can be seen to give 
rise to negative values of the average correlation function. \\
\noi We initially investigate the asymptotic behaviour of the $h$ function 
for such a system, and obtain a fitting function for $h(\delta)$ when 
$\delta < 0$. These results will be used to argue for a form of $h(\xib)$, 
which can be then used to obtain an NSR between $\xib$ and $\xib_L$, in the 
negative $\xib$ regime.\\

\noi The general equation for the evolution of a spherical shell 
which encloses a mass $M$ can be written as 
\beq
\label{reqn}
\frac{d^2R}{dt^2} = -\frac{GM}{R^2} \;\; ,
\eeq
where $R(t)$ is the proper radius of the shell; the mass $M$ is a 
constant, in the absence of shell crossing. The first integral 
of equation (\ref{reqn}) is 
\beq
\frac{1}{2}\left(\frac{dR}{dt}\right)^2 = \frac{GM}{R} + E \; .
\eeq
The constant $E$ gives the total energy of the shell; clearly, 
$E > 0$ for a shell with radius, $R < R_2$ while $E = 0$ for a shell with 
radius, $R > R_2$. We will set up the initial conditions by assuming 
that the shell has radius $R_i$ at the initial time, $t_i$ and moves
with the general background expansion at this time, with zero 
peculiar velocity. The total energy of the shell is then given 
by \cite{pad96b}
\beq
E = -K_i\delta_i \; \; ,
\eeq
where $K_i = (1/2)H_i^2R_i^2$ is the initial kinetic energy, $\delta_i$ 
is the initial density contrast and $H_i$ is the Hubble parameter at 
the initial time; note that $\delta_i < 0$, so that $E > 0$.
\\
\noi The solution to equation (\ref{reqn}), with these initial conditions, 
can be written in the parametric form 
\begin{eqnarray}
\label{rsoln}
R &=& \frac{GM}{2E}\left[{\rm{cosh}}\;\theta - 1\right] \\
\label{tsoln}
t&=&\frac{GM}{(2E)^{3/2}}\left[{\rm{sinh}}\;\theta - \theta\right] \; .
\end{eqnarray}
\noi The peculiar velocity, $v$,  defined as $v = a{\dot x}$, is \\
\beq 
v \:=\: \frac{d}{dt}\left[ax\right] - {\dot a}x\: =\: {\dot R} - HR \; .
\eeq
\noi This implies that
\beq 
\label{heqn}
h \equiv -\frac{v}{HR} = 1 - \frac{\dot R}{RH} \; .
\eeq
\noi For an $\Omega = 1$, matter dominated Universe, $H = (2/3t)$; substituting for 
$H$ and ${\dot R}$ in equation (\ref{heqn}) yields
\beq
h = \left[ 1 - \frac{3t}{2R} \sqrt{2\left(\frac{GM}{R} + E\right)}\right] \; .
\eeq
\noi We now use equations (\ref{rsoln}) and (\ref{tsoln}) to rewrite the 
above in terms of the parameter $\theta$ as 
\beq
\label{htheta}
h = 1 - \frac{3}{2}\frac{\rm{cosh}\;(\theta/2)}{\rm{sinh}\;(\theta/2)}\left[
\frac{\rm{sinh}\;\theta - \theta}{\left({\rm{cosh}}\;\theta - 1\right)^{3/2}}\right] \; .
\eeq
\noi Further, the density contrast $\delta$ is given by 
\beq
1 + \delta = \frac{9GMt^2}{2R^3} \; .
\eeq
\noi Again replacing for $R$ and $t$ in terms of $\theta$ gives 
\beq
\label{deltheta}
\delta = \frac{9}{2}\frac{({\rm{sinh}}\;\theta - \theta)^2}
{\left({\rm{cosh}}\;\theta - 1\right)^3} - 1 \; .
\eeq
\noi In the limit $\theta \rightarrow \infty$, $\delta \rightarrow -1$ and 
$h \rightarrow -1/2$. We thus have the interesting result that $h \rightarrow 
-1/2$ as $\delta \rightarrow -1$.\\
\begin{figure}
\centering
\psfig{file=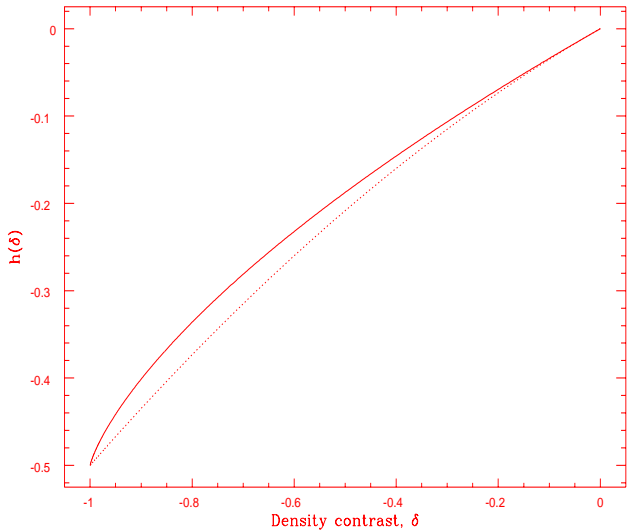,width=3.5truein,height=3.0truein,angle=-90}
\caption{The figure shows the $h$ function, plotted versus density contrast, 
$\delta$, for a compensated void (solid line). The dotted line indicates a fit of 
the form $h(\delta) = a\delta + b\delta^2$, with $a = 1/3$, $b = -1/6$.}
\label{fig1}
\end{figure}
\noi A good fit to the function $h(\delta)$, using equations (\ref{htheta}) and 
(\ref{deltheta}), is given by $h(\delta) = \delta/3 - \delta^2/6$ (see figure 
(\ref{fig1})). The fit satisfies both the linear limit ($h(\delta) \sim \delta/3$, 
for $|{\delta}| \ll 1$), as well as tends to -1/2 as $\delta \rightarrow -1$. The 
percentage error on the fit is less than 10\%. \\

\noi The preceding analysis of an individual compensated void has yielded a fitting 
function for $h(\delta)$; we, however, require $h$ as a function of $\xib$, 
 to find the NSR for the negative $\xib$ regime. There is -- in general -- no simple 
relation between the density contrast $\delta$ of a single lump and the correlation 
function $\bar\xi$. (Purists will even consider the question of relating $\bar\xi$ 
and $\delta$ meaningless.) Despite this, relations obtained for single spherical 
lumps have proved to be generalizable to a description of statistical quantities 
like $\bar \xi$ \cite{pad96,pad97,mowhite,sheth,ekp}. In this spirit, 
and motivated by the relation between $h$ and $\delta$ being 
quadratic, we will attempt an ansatz in which $h$ is a quadratic function of 
$\xib$. In the linear regime, however, $h = (2/3)\xib$; this result arises from 
straightforward application of perturbation theory \cite{pad96} 
and is independent of the sign of $\xib$. Next, $\xib$ and $\delta$ are both bounded 
from below by -1 and $\xib \propto \delta$ in the non-linear regime \cite{ekp}. 
It thus seems reasonable that $\xib \approx -1$ when $\delta = -1$; this immediately 
implies that $h \sim -1/2$ when $\xib \approx -1$.  We will hence choose 
\beq
h(\xib) = \frac{2}{3}\xib + \frac{1}{6}\xib^2 \; \; ,
\eeq
\noi similar to the form of $h(\delta)$ and satisfying the condition $h \rightarrow 
-1/2$ as $\xib \rightarrow -1$. This will be used in equation (\ref{xilxi}) to 
obtain an NSR for the negative $\xib$ regime.\\
\end{document}